\documentclass[preprint,preprintnumbers, prd, floatfix,  superscriptaddress,
nofootinbib] {revtex4}
\usepackage{graphicx}
\usepackage{epsfig}
\usepackage{bm}
\usepackage{amssymb}
\usepackage{float}
\usepackage{amsmath}
\usepackage{dcolumn}
\usepackage{cancel}
\usepackage[colorlinks]{hyperref}
\usepackage[usenames,dvipsnames]{color}
\usepackage{epstopdf}
\hypersetup{
     breaklinks=true,
    pdfstartview={FitH},    
    colorlinks=true,       
    linkcolor=blue,          
    citecolor=red,        
    filecolor=magenta,      
    urlcolor=blue,           
    anchorcolor=green,      
    linktocpage=true
}

\newcommand{\Mpl}{M_{\textrm{Pl}}}
\renewcommand{\(}{\left(}
\renewcommand{\)}{\right)}

\def\al{\alpha}
\def\bet{\beta}

\def\Om{\Omega}

\def\S{\mathcal{S}}

\def\C{\mathcal{C}}

\def\doi{http://doi.org}
\def\arxiv{http://arxiv.org/abs}
 
 \def\e{\mathrm{e}}
\def\r{\mathrm{r}}

\def\m{\mathrm{m}}

\def\d{\mathrm{d}}

\begin{document}

 \title{Observational constraints on varying neutrino-mass cosmology}

 \author{Chao-Qiang Geng}
\email{geng@phys.nthu.edu.tw}
\affiliation{Chongqing University of Posts \& Telecommunications, Chongqing, 400065, 
China}
 \affiliation{National Center for Theoretical Sciences, Hsinchu,
Taiwan 300}
\affiliation{Department of Physics, National Tsing Hua University,
Hsinchu, Taiwan 300}

\author{Chung-Chi Lee}
\email{g9522545@oz.nthu.edu.tw}
 \affiliation{National Center for Theoretical Sciences, Hsinchu,
Taiwan 300}

\author{R.~Myrzakulov}
\email{rmyrzakulov@gmail.com} 
\affiliation{ Eurasian  International
Center for Theoretical Physics, Eurasian National University, Astana
010008, Kazakhstan}

\author{M.~Sami}
\email{sami@iucaa.ernet.in} \affiliation{Centre for Theoretical
Physics, Jamia Millia Islamia, New Delhi-110025, India}

\author{Emmanuel~N.~Saridakis}
\email{Emmanuel\_Saridakis@baylor.edu}
\affiliation{Physics Division,
National Technical University of Athens, 15780 Zografou Campus,
Athens, Greece}
 \affiliation{Instituto de F\'{\i}sica,
Pontificia Universidad de Cat\'olica de Valpara\'{\i}so, Casilla
4950, Valpara\'{\i}so, Chile}

\begin{abstract}
 We consider generic models of quintessence and we investigate the influence of massive 
neutrino 
matter with field-dependent masses on the matter power spectrum. In case of minimally 
coupled 
neutrino matter, we examine the effect in tracker models with inverse power-law and 
double 
exponential potentials. We present detailed investigations for the scaling field with a 
steep 
exponential potential, non-minimally coupled to massive neutrino matter, and we derive 
constraints on field-dependent neutrino masses from the  observational data.
\end{abstract}


\maketitle

\section{Introduction}

It is remarkable that both early and late time completions of the standard model
of the universe include phases of accelerated expansion. 
Inflation~\citep{Starobinsky:1980te, Guth:1980zm,Linde:1981mu}
not only successfully addresses early
time inconsistencies of the hot big bang, such as the flatness and horizon problems, but 
it also provides the mechanism of primordial perturbations needed for structure formation 
in the universe. On the other hand, the age crisis~\citep{Krauss:1995yb} in the standard
cosmological model again asks for a late-time cosmic acceleration for its resolution in
the standard lore. This phenomenon was directly confirmed by supernovae observations in
1998~\citep{Riess:1998cb,Perlmutter2}, and it 
was indirectly supported by other probes
thereafter~\citep{Spergel:2003cb, Lange:2000iq, Ade:2013zuv}. Such a confirmation for
inflation is still awaited.

The late-time cosmic acceleration can be caused by the cosmological
constant~\citep{Peebles:2002gy}, with an energy scale of 
$O(10^{-3})$~eV, or by a slowly-rolling scalar field (quintessence) with
the mass of the order of $H_0\sim 
10^{-33}$~eV~\citep{Fujii:1982ms, Ford:1987de, Ratra:1987rm, Wetterich:1987fk}. Both
options are plagued with deep theoretical problems. Concerning the
cosmological constant, it is an ``unnatural'' parameter~\citep{Hooft}
of Einstein theory, and its small numerical value at the classical
level gets destabilized by quantum correction$-$vacuum energy.
Concerning the fundamental scalar field, it faces the same problem
of ``naturalness''. In Minkowski space time, vacuum energy can be
safely ignored by choosing normal ordering in quantum field theory.
In this case, it neither  influences the dynamics nor there is a
known way to measure it in local experiments. However, in curved
space time, vacuum energy adds to the energy momentum of matter and
it contributes to the dynamics of spacetime geometry. In the
important case of Freidmann-Robertson-Walker (FRW) cosmology,
geometry is conformally equivalent to Minkowski spacetime and thus
one might naively think that  it has solved the cosmological
constant problem. In this case, however, one is left with a  scalar
field non-minimally coupled to matter in flat spacetime~\citep{Fujii_Maeda, Faraoni}. 
Hence, the cosmological problem translates
into an equivalent problem of naturalness of the scalar field.

Leaving the aforementioned problem aside, if we opt for quintessence
then a specific behavior of the scalar-field dynamics is required,
in order to keep the thermal history of the universe intact.
Moreover, if quintessence has originated from inflation in the early
universe in an attempt to attribute both accelerating phases in the
same scalar field (quintessential 
inflation)~\citep{Peebles:1999fz, Copeland:2000hn, Sahni:2001qp, Sami:2004xk, 
Hossain:2014zma}, 
then the scaling behavior of
the scalar field is required in order to comply with the
nucleosynthesis constraint. The exit from the scaling regime to the
late-time acceleration can be caused by invoking the non-minimal
coupling with massive neutrino 
matter~\citep{Wetterich:2013jsa, Hossain:2014coa, Hossain:2014xha}. 
It is really mysterious that the neutrino mass is around the mass scale
associated with dark energy. Once the non-minimal coupling to
neutrino matter is invoked, the mass of the neutrino gets connected
to the minimum of the effective potential and thereby to dark
 energy~\citep{Hossain:2014zma}. The neutrino mass modifies the 
cosmological
evolution at both background and perturbation levels~\cite{Mota:2008nj, Ayaita:2014una}. 
Amongst
others, the neutrino mass shifts the time of matter-radiation
equality, and the free-streaming massive neutrino suppresses the
growth of matter density perturbation within the free-streaming
length scale~\citep{Lesgourgues:2006nd, Geng:2014yoa}. It is well
known that particle-physics experiments give rise to constraints on
neutrino masses. On the other hand, cosmology provides an
independent probe, which is however sensitive to the new degree(s)
of freedom over and above the standard model of particle physics.
The massive neutrino matter crucially affects the matter power
spectrum, and thus it can give rise to constraints on neutrino
masses. In particular, the matter-density power spectrum is damped
on small scales by massive neutrinos.

Having in mind the above discussion, it is both interesting and
necessary to investigate the observational constraints on varying
neutrino-mass cosmology. Although observational constraints analysis on varying-mass 
neutrino cosmology has been performed in the past (see for instance~\citep{Fardon:2003eh, 
Peccei:2004sz, Brookfield:2005bz, Amendola:2007yx, Wintergerst:2009fh, LaVacca:2012ir})  
the crucial new 
feature of our investigation is the 
incorporation 
of the non-minimal coupling. In particular, we shall investigate
quintessence models based upon the tracker field with massive
neutrino matter minimally as well as non-minimally coupled to the
field. Moreover, another novel feature is the use of the recently released Planck and 
SDSS 
data. The plan of the manuscript is the following: In
section~\ref{sec:2}, we consider dark energy scenarios based upon
scalar field models, with inverse power law and double exponential
potentials with minimally coupled neutrino matter.
Section~\ref{model} is devoted to the  dynamics of a scaling field
coupled to massive neutrino matter. In section~\ref{sec:4} we present our results on  the
observational constraints on the above models. Finally, we give our conclusions in
section~\ref{sec:conclusion}.

\section{Quintessence  minimally coupled to matter}
\label{sec:2}

The quintessence scenario is one of the main paradigms for the
description of the late-time acceleration~\citep{Copeland:2006wr}. In
this set up, along with cold dark matter and baryonic component, we
shall also be interested in considering  massive neutrino matter.
As mentioned above, this class of models can give rise to
quintessential inflation, such that the dark energy scale  emerges
naturally in the scenario.

Before proceeding to non-minimal  coupling with
neutrino matter, we first briefly consider standard quintessence
with minimally coupled massive neutrino matter.
In order to set
the notations, let us consider the following standard action
\begin{eqnarray}
\mathcal{S} = \int d^4x \sqrt{-g} \left[ -\frac{\Mpl^2}{2}R + \frac{1}{2}\partial^{\mu}
\phi \partial_{\mu} \phi + V(\phi) \right] + S_m+ S_r \,,
\label{action00}
\end{eqnarray}
where $S_{m,\ r}$ denotes the action of ordinary matter including
non-relativistic and relativistic one. Additionally, in order to
describe the background cosmological evolution, we consider the
spatially flat FRW geometry, in which case the action
(\ref{action00})  leads to the following evolution equations
\begin{eqnarray}
\label{eq:Friedmann-1}
&&3H^2 \Mpl^2 = \rho_m +\rho_r + \rho_{\phi} \,, \\
\label{eq:Friedmann-2} &&\left( 2\dot{H}+3H^2\right)\Mpl^2=-\left(
p_m+p_r + p_{\phi} \right) \,,\\
&& \ddot{\phi} + 3H\dot{\phi} +\frac{dV}{d\phi} =0 \,,
\label{eq:quinteom}
\end{eqnarray}
where $\rho_\phi$ and $p_\phi$ denote the energy density and
pressure of the quintessence field.
 In this case as usual, the dark energy sector is attributed to $\phi$ and the
dark-energy equation-of-state parameter reads as $w_{DE}=w_\phi \equiv p_{\phi}/ 
\rho_{\phi}$.

As mentioned in the introduction, specific features of scalar field
dynamics such as tracker solutions are of interest in cosmology. In
this case, once the conditions at the present epoch are set
properly, cosmic evolution is not sensitive to the initial conditions. In what
follows, we shall try to alleviate the problem associated with the
said choice, for model parameters attributed to tracker solutions.

To make the point, we shall consider tracker solutions in the models
with inverse power-law and double-exponential type potentials,
though the underlying features are common to any tracker model.
Let us begin with the inverse power-law type potential~\citep{Ratra:1987rm, 
Kneller:2003xg, Abramo:2003cp, Saridakis:2009pj}
\begin{eqnarray}
\label{eq:potphi4}
V(\phi) = V_0 \left( \frac{\phi}{\Mpl} \right)^{-n}\,,
\end{eqnarray}
where $V_0$ is a mass dimension-four constant. In this case, the
slope of the potential is given by $\lambda_\phi = -\Mpl V^{\prime}/V = n/\phi$, which
is large at early times and diminishes at late times when $\phi$
acquires large values. Consequently, the field might mimic the
background ({\it scaling solution}) at early epochs and could exit
to slow roll at late times, giving rise to de Sitter attractor {\it a
la a tracker solution}. As explained in the Appendix, it is
difficult to achieve tracker in this case for generic values of
model parameters. For instance, for $V_0\sim M_p$, one requires much
larger value of $n$ (see Ref.~\citep{Sahni:2001qp} for 
details). It
turns out that it is much easier to realize tracker in a model with the
double exponential potential.

Let us  consider the  potential of the form~\citep{Barreiro:1999zs, Gonzalez:2006cj, 
Gonzalez:2007hw}
\begin{eqnarray}
\label{eq:potcosh}
V(\phi)=V_0 \left( e^{-c_1 \frac{\phi}{\Mpl}} + e^{c_2 \frac{\phi}{\Mpl}} \right) \,,
\end{eqnarray}
where $V_0$, $c_1$ and $c_2$ are constants. For $\phi_i \gg 0$ at
the initial time ($N \equiv \ln a \rightarrow -\infty$), the second term
  in (\ref{eq:potcosh}) can give rise to scaling regime for
generic values of $c_2>9$ consistent with nucleosynthesis constraints of 
$\Omega_{\phi}=3(1+w_b)/c_2^2<0.045$~\citep{Bean:2002sm, Amendola:2007yx}, where $w_b$ is 
the equation-of-state of the 
background fluid, i.e., $w_b=1/3$ and $0$ in radiation and matter dominated epochs, 
respectively. 
On the other hand,  
for late times, when $\phi$ approaches the origin,
(\ref{eq:potcosh}) exhibits a minimum with $V_{min}\sim V_0$ for
$c_2\sim c_1$. The latter can suit our requirement if we choose
$V_0\sim \rho_{cr}$.

\begin{figure}
\centering
\includegraphics[width=.45\linewidth]{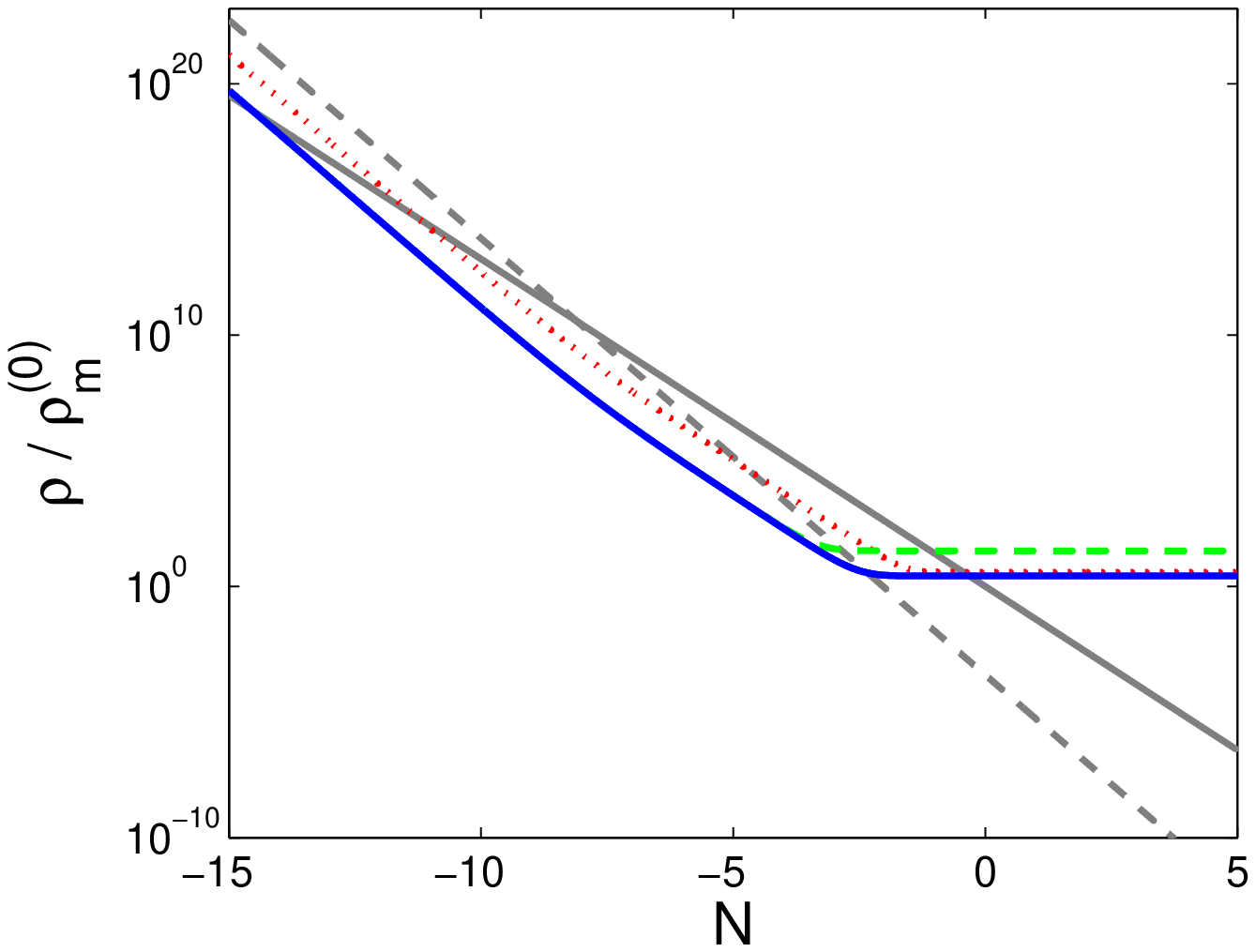}
\includegraphics[width=.45 \linewidth]{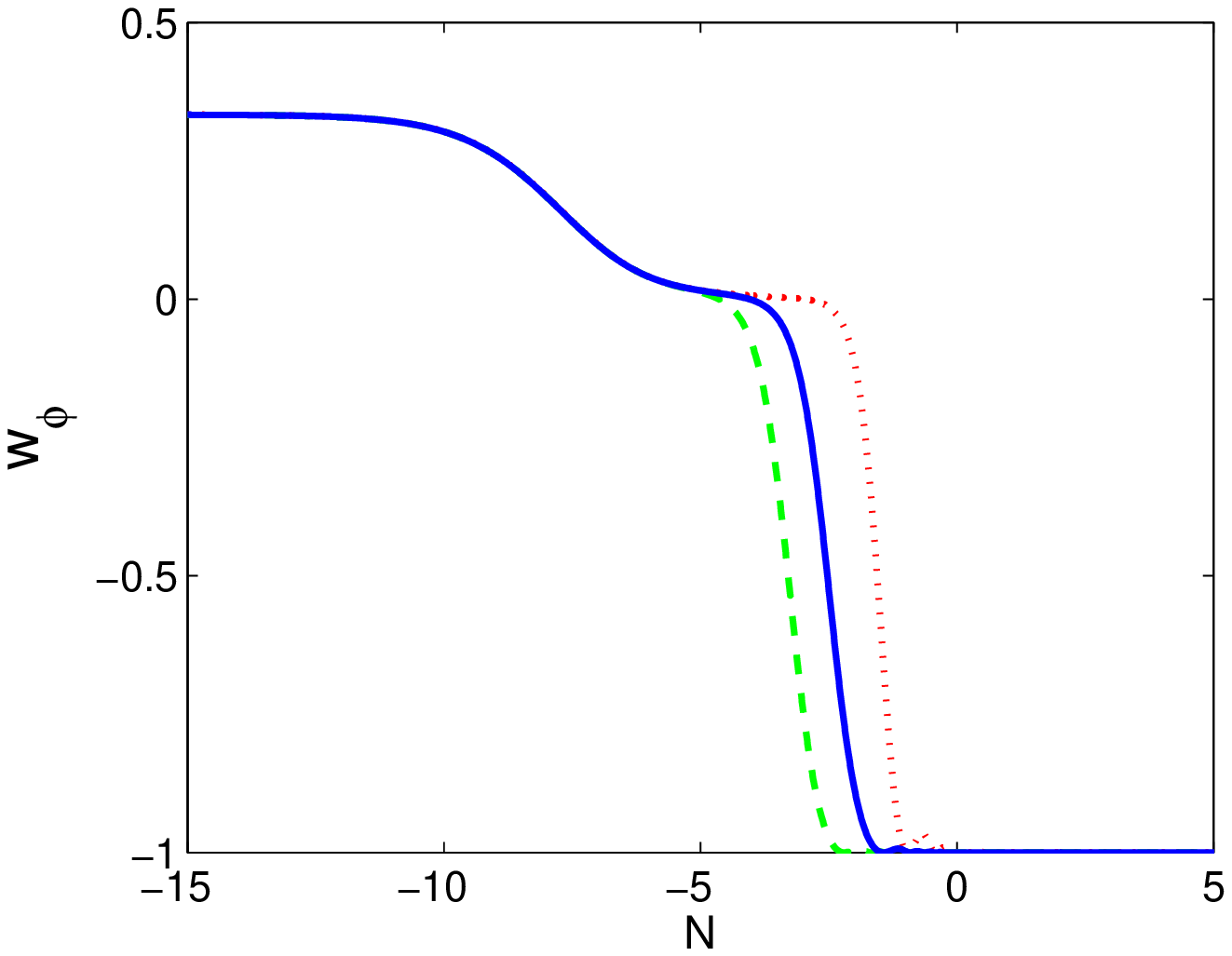}
\caption{ {\it{Left: Evolutions of  various energy densities,
normalized to matter energy density $\rho_{m}^{(0)}$ at present
($a_0=1$), as functions of $N \equiv \ln a$, for the minimally
coupled scenario with the double-exponential potential $V=V_0 (
e^{-c_1 \phi/\Mpl} + e^{c_2 \phi/\Mpl} )$: $\rho_r$ is the
gray-dashed curve, $\rho_m$ is the gray-solid curve, and
$\rho_{\phi}$ is plotted for three parameter choices, namely ($c_1,
c_2, V_0/\rho_{m}^{(0)}$)= $(4, 50, 2)$ (blue-solid),  ($c_1, c_2,
V_0/\rho_{m}^{(0)}$)= $(4, 50, 20)$ (green-dashed) and ($c_1, c_2,
V_0/\rho_{m}^{(0)}$)= $(4, 10, 2)$ (red-dotted). 
Right: The
corresponding evolution of the quintessence equation-of-state
parameter $w_\phi$. We have used $\rho_r^{(0)}/ \rho_m^{(0)} = 3
\times 10^{-4}$ as the boundary condition.}}} \label{fig3}
\end{figure}

\begin{figure}
\centering
\includegraphics[width=.45 \linewidth]{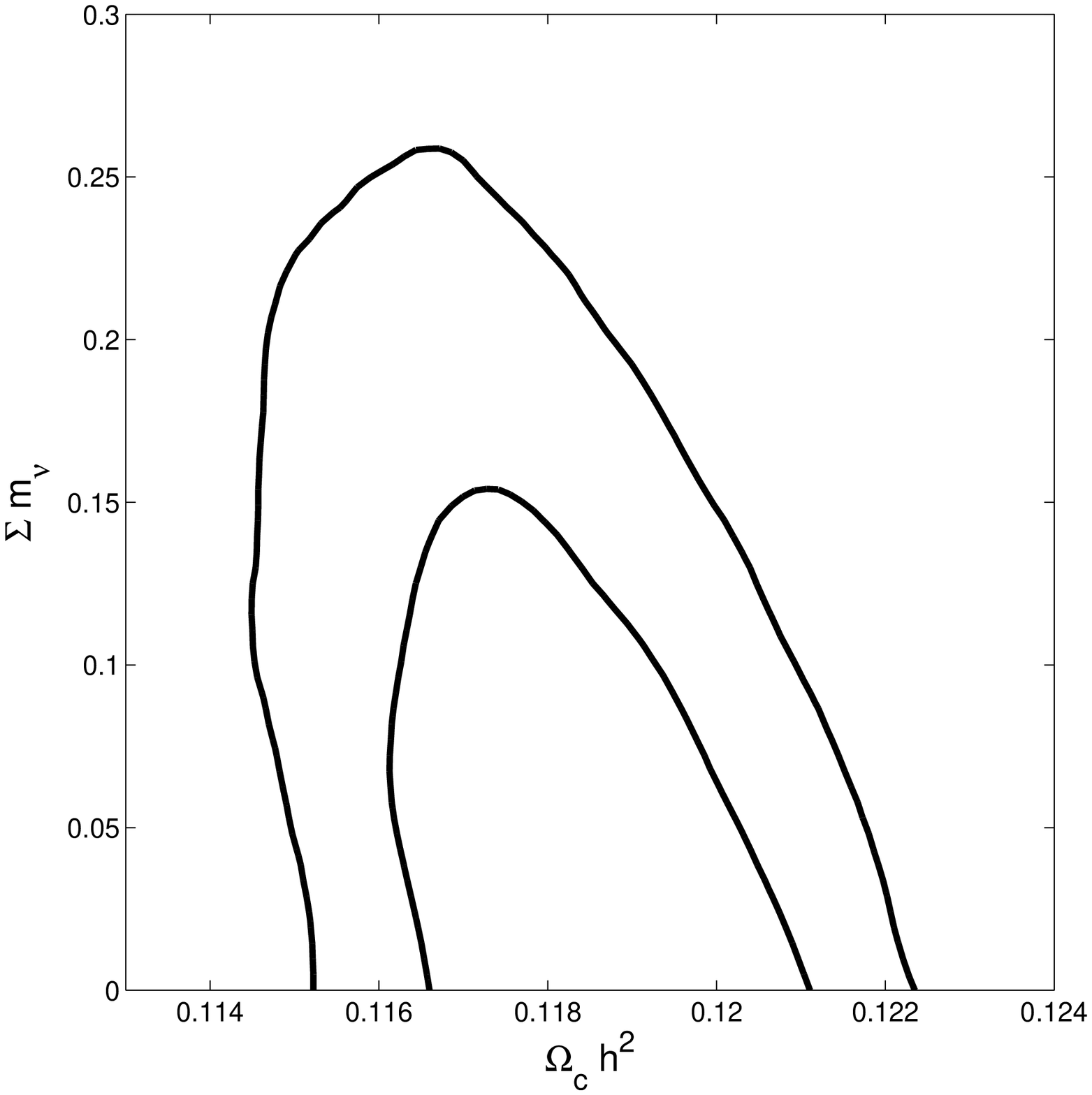}
\includegraphics[width=.45 \linewidth]{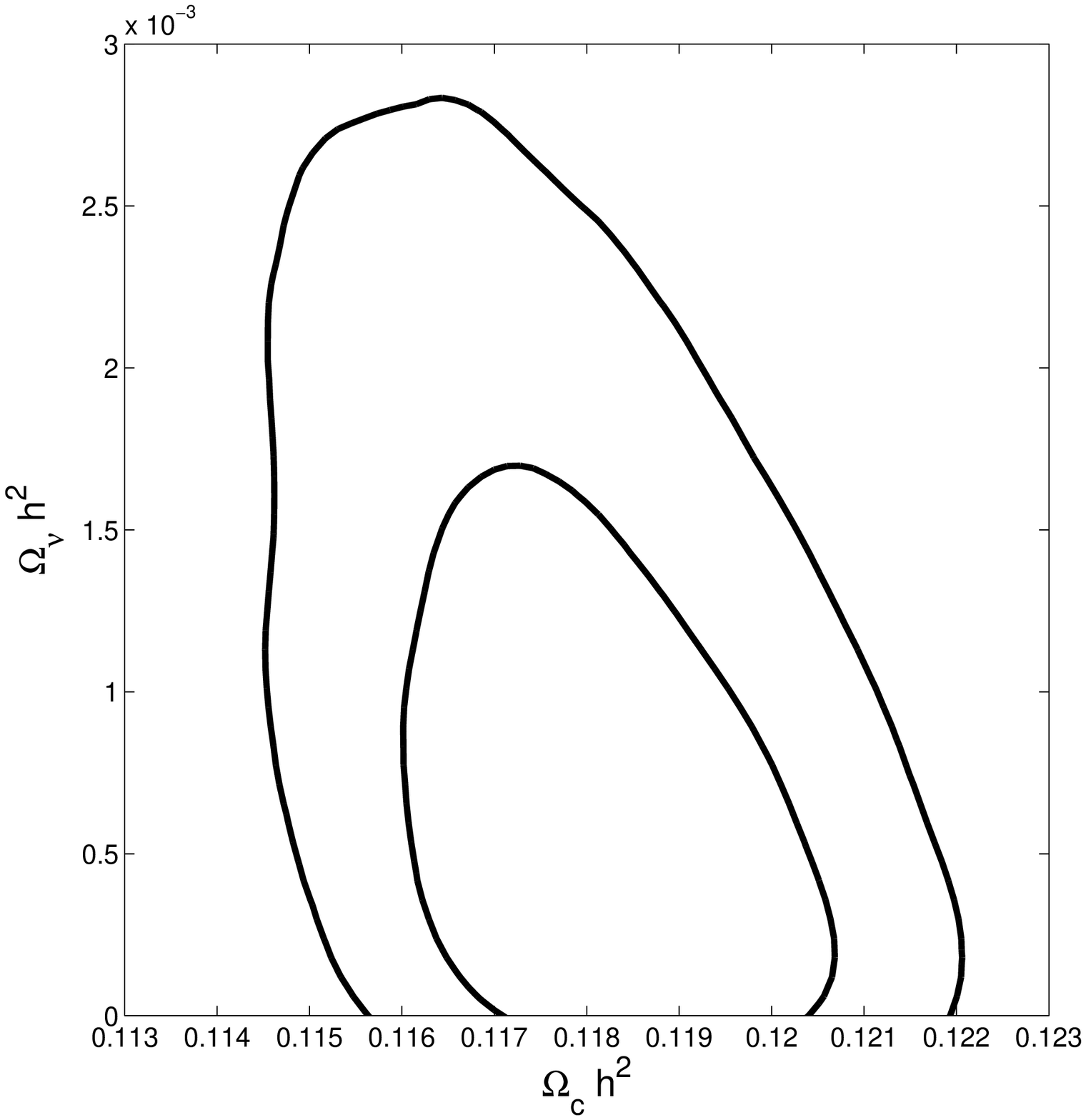}
\caption{{\it{Likelihood contours of the mass sum of the three neutrino species $\Sigma 
m_{\nu}$ 
and energy density ratio $\Omega_{\nu}h^2$
 versus the CDM physical density $\Omega_c h^2$, for the
minimally coupled scenario with the effective potential $V(\phi)
=V_0 \left( e^{-c_1 \phi/\Mpl}+e^{c_2 \phi/\Mpl} \right)$, where the inner and outer 
curves 
represent $1$ and $2\sigma$
confidence levels, respectively.
 }}}
\label{fig6}
\end{figure}

In order to present the aforementioned behavior in a more
transparent way, we numerically evolve the evolution equations and
we depict the corresponding evolution in Fig.~\ref{fig3}. Indeed, in the
upper graph of Fig.~\ref{fig3}, we depict the corresponding
evolutions for the energy densities of radiation ($\rho_r$), matter
($\rho_m$) and quintessence field ($\rho_{\phi}$), normalized to the
matter energy density $\rho_{m}^{(0)}$ at present, as functions of
$N \equiv \ln a$, for three choices of $V_0$, $c_1$ and $c_2$. In
the lower graph of Fig.~\ref{fig3}, we depict the corresponding
evolution for the quintessence equation-of-state parameter $w_\phi$.
Because the cosmic evolution is insensitive to the initial conditions in the exponential 
potential, 
we can choose initial values in a broad parameter space. In the calculation, 
$\phi/\Mpl=30/c_2$ and 
$\dot{\phi}=0$ are set at $N = -20$.
As we observe, the density ratio $\rho_{\phi}/\rho_{m}^{(0)}$,
depending on $c_2$ in the early universe, does not change in
radiation and matter dominated eras. Finally, the behavior of
$\phi$-field at early times, can be estimated by combining
(\ref{eq:quinteom}) and (\ref{eq:potcosh}), leading to $w_{\phi} =
w_b$. Hence, as the scale of $\rho_{\phi}$ is close to $V_0$, the
quintessence field evolves into the $\Lambda$CDM-like stage, where
$w_{\phi} \rightarrow -1$. Note that because the
late-time behavior of the double exponential potential is almost the
same as
 that of the $\Lambda$CDM model, this scenario is hardly  distinguished from $\Lambda$CDM.
 We perform the CosmoMC package~\citep{Lewis:1999bs, Lewis:2002ah} to extract the 
observational 
constraints and present the results of these two scenarios 
with one massive neutrino along with the other two being massless
in Fig.~\ref{fig6} and Table.~\ref{table3}.
The obervational dataset will be introduced in Sec.~\ref{sec:4}.
\begin{table*}
\begin{center}
\caption{List of priors for parameters and allowed regions with
$95\%$ C.L., and
$V(\phi) =V_0 \left( e^{-c_1 \phi/\Mpl}+e^{c_2\phi/\Mpl}
\right)$ and 
$c_1=20$.  }
\begin{tabular}{lccc} \hline\hline
Parameter & Prior & Our result ($95\%$ C.L.) & $\Lambda$CDM ($95\%$ C.L.)
\\ \hline
Baryon density & $0.5 <100\Omega_bh^2<10$ & $ 100 \Omega_bh^2 =   2.20 ^{+0.04}_{- 0.05}$ 
&
$ 100 \Omega_bh^2 =  2.22^{+0.04}_{-0.06}$
\\ \hline
CDM density & $10^{-3}<\Omega_ch^2<0.99$ & $ \Omega_ch^2 =  0.118 \pm 0.003$ & $
\Omega_ch^2
=  0.117^{+0.004}_{-0.002}$
\\ \hline
Neutrino mass & $0<\Sigma m_{\nu} < 1 $ eV & $\Sigma m_{\nu} < 0.200$ eV
& $\Sigma m_{\nu} < 0.198$ eV
\\ \hline
Spectral index & $ 0.9 < n_s < 1.1$ & $ n_s = 0.961 \pm 0.011$ &
$ n_s = 0.963^{+0.010}_{-0.011} $
\\ \hline
Potential  & $ 10 < c_2 < 40$ & $c_2 > 24.6$ & $-$
\\ \hline
\end{tabular}
\label{table3}
\end{center}
\end{table*}

It is clear from the above that the power-law potentials
inevitably lead to a dark-energy dominated epoch, and this
$\phi$-dominated era occurs when $\rho_{\phi}$ is of the same order
as $\rho_m$. However, for generic values of model parameters, the
scenario is severely constrained  by observations.
 On the other
hand, the double exponential potential leads to a different
behavior, in which a natural evolution in the radiation, matter
and late-time acceleration regimes can be realized,
provided that we choose appropriately the model parameters. In
particular, the appearance of the dark energy dominated era
crucially depends upon the scale of the potential $V_0$, which should
be chosen around $\rho_{cr}$.  The latter brings back the original
problem associated with the cosmological constant.

In this section, we have performed the analysis in tracker models that have generic 
features related to the insensitivity of evolution to initial conditions. In this case, 
our results are close to the $\Lambda$CDM paradigm \citep{Ade:2013zuv}. This is not 
surprising since we are dealing with the slowly-rolling scalar field, which mimics the 
cosmological constant at late times. Therefore, we do not have significant improvements 
compared to the predictions of the $\Lambda$CDM scenario. It is clear that a successful 
model of quintessence should maintain the scaling behavior, with an additional mechanism 
of exit to dark energy that could avoid the above disadvantage. This mechanism is
provided by the non-minimal coupling between quintessence and massive neutrino matter. In 
the section to follow, we construct and analyze a model with the above features.

\section{Non-minimally coupled massive
neutrino matter and  graceful exit   to dark energy} \label{model}

As mentioned above, one should look for a scenario in which the
scale of dark energy emerges somewhat naturally. In this section we
briefly review a model  in which a key mechanism is the nonminimal
coupling of the scalar field with  varying-mass neutrinos~\citep{Hossain:2014zma}. In 
this 
case, $V_0$ gets linked to the energy
density of massive neutrino matter.

The relevant action in the Einstein frame is written as
  \begin{eqnarray}
&\mathcal{S} = \int d^4x \sqrt{-g}\bigg[-\frac{\Mpl^2}{2}R+
\frac{1}{2}\partial^\mu\phi\partial_\mu \phi+V(\phi)
\bigg]\nonumber\\
&\!\!\!\!\!\!\! \!\!\!\!\!\!\!+\S_\m+\S_\r+\S_\nu\(\C^2g_{\al\bet},\
\Psi_\nu\) \, .
 \label{action1}
\end{eqnarray}
In the above expression, we have introduced a scalar field $\phi$
with $V(\phi)$ its potential, as well as  matter and radiation
sectors corresponding to perfect fluids. Note that due to the
nonminimal curvature-neutrino coupling in the Jordan frame, we
obtain a varying-mass neutrino sector in the Einstein frame,
quantified by the coefficient~\citep{Hossain:2014zma}
\begin{equation}
 \C^2=A^2 \e^{2\beta\phi/\Mpl} \, .
\end{equation}

Hence, varying the action (\ref{action1}) with respect to the metric
and specializing to the FRW case, we obtain the evolution equations
\begin{eqnarray}
 &&3H^2\Mpl^2 = \frac{1}{2}\dot\phi^2+V(\phi)+\rho_\m+\rho_\r+\rho_\nu \, ,
 \label{FR1} \\
 &&\(2\dot H+3H^2\)\Mpl^2 = -\frac{1}{2}\dot\phi^2+V(\phi)-p_m-p_\r-p_\nu  , \, ~~~~~~
 \label{FR2}
\end{eqnarray}
where $\rho_i$ are the energy densities of the corresponding
sectors, and $p_i$ their pressures. Additionally, varying the action
(\ref{action1}) with respect to the scalar field, we derive its
equation of motion, which reads
\begin{equation}
\label{phieq}
 \ddot\phi+3H\dot\phi+\frac{\d V}{\d \phi}=-\frac{\beta}{\Mpl}(\rho_\nu-3p_\nu) \,,
\end{equation}
and thus the evolution equation for the neutrinos becomes
\begin{eqnarray}
\label{neutrinoeq}
 \dot\rho_\nu+3H(\rho_\nu+p_\nu)=\frac{\beta} {\Mpl}\dot\phi
(\rho_\nu-3p_\nu) \, .
\end{eqnarray}
Moreover, concerning radiation and matter sectors, we have the standard evolution.
Additionally, for the neutrino sector we have
\begin{eqnarray}
 m_{\nu,\rm eff}(\phi)=m_{\nu,0}e^{\beta\phi/\Mpl} \, ,
 \label{eq:mnu}
\end{eqnarray}
and thus the neutrino pressure $p_{\nu}$ behaves as radiation during
the early times, while it behaves as non-relativistic matter during
late times~\citep{Hossain:2014xha} where non-minimal coupling builds
up to induce non-zero neutrino masses. In this case, at late times, one obtains an 
effective dark
energy sector, consisting of the scalar field as well as of the
varying-mass neutrinos. In this case, the energy density and pressure
of dark energy are given by
\begin{eqnarray}
\label{rhoDE}
 &&\rho_{\rm
DE}\equiv\rho_\nu+\rho_\phi=\rho_\nu+\frac{1}{2}\dot\phi^2+V(\phi),
 \\
 &&p_{\rm DE}\equiv p_\nu+p_\phi=p_\nu+\frac{1}{2}\dot\phi^2-V(\phi),
 \label{pDE}
\end{eqnarray}
which obey the standard
 continuity equation.
Finally, as usual  it proves to be convenient to  introduce the
dimensionless density parameters for radiation, matter, neutrinos
and scalar field as
\begin{eqnarray}
 \Omega_i &=& \frac{\rho_i}{3H^2\Mpl^2} \, ,~~~(i=m\,,\ r\,,\ \nu\,,\ \phi)
 \label{Omega_m}
\end{eqnarray}
 respectively, and therefore from (\ref{rhoDE}) we obtain
 \begin{eqnarray}
 \Om_{\rm DE}=\Om_\phi+\Om_\nu \, .
 \label{eq:density_DE0}
\end{eqnarray}

In order to obtain a viable  model, we have to consider a
specific ansatz for the potential $V(\phi)$ which is dictated by the
requirement of nucleosynthesis. Hence, in order to illustrate the
neutrino matter effect we consider the scaling potential, namely
\begin{equation}
V(\phi) = V_0 e^{-\alpha \phi/\Mpl} \,.
\end{equation}
The exponential potential in the non-minimally coupled quintessence model has been 
investigated with $\alpha<\sqrt{2}$~\citep{Brookfield:2005bz, LaVacca:2012ir}, satisfying 
the slow-roll potential condition. 
On the other hand, the scaling potential, $\alpha \gg 2$, has also been discussed to 
address the 
late-time acceleration occurring due to the effective potential of the non-minimally 
coupled 
neutrino matter~\citep{Amendola:2007yx}.
Our work follows the later ``naturalness'' scaling potential with current observations.
Thus, the scalar field equation of motion (\ref{phieq}) becomes
\begin{eqnarray}
\label{eq:fieldeq}
\ddot{\phi} + 3H \dot{\phi} + V_{eff, \phi}=0 \,,
\end{eqnarray}
where
\begin{eqnarray}
\label{eq:poteff}
V_{eff, \phi}  = \frac{d V}{d \phi}
+ \frac{\beta}{\Mpl} (\rho_{\nu} - 3 p_{\nu})
\label{Veff}
\end{eqnarray}
is the derivative of the effective potential in which the scalar
field moves,  constituted of the original potential as well as of extra
terms  contributed by massive neutrino matter coupled to scalar
field. The above cosmological scenario proves to exhibit a very
interesting behavior, since this effective potential develops a
minimum. We note that the latter provides an interesting mechanism
for the unified description of inflation and late-time 
acceleration~\citep{Hossain:2014xha,Hossain:2014ova}. 
Indeed, we easily check that
the minimization of $V_{eff}$ requires,
\begin{eqnarray}
\label{phivmin}
 \phi_{min}=\frac{\Mpl}{\alpha}\ln\left(\frac{V_0}{\gamma
 \rho_\nu}\right);~~~\beta\equiv \alpha\gamma
\end{eqnarray}
which leads to
\begin{eqnarray}
\label{eq:de-approach}
V_{eff}^{min}=\left(1+\gamma\right)\rho_\nu(\phi_{min})\,,
\end{eqnarray}
where we have ignored the neutrino pressure at late times.
 We observe two features from (\ref{eq:de-approach}). Firstly, it
is clear that $V_{eff}^{min}$ is not very sensitive to $V_0$ and
that the minimum of the effective potential is mainly defined by the
numerical value of neutrino matter density at the present epoch.
Secondly, the late-time dark energy density depends on the model
parameters $\gamma$. It is interesting to note that
$\omega_{DE}\equiv-1-2\dot{H}/3H^2=-\gamma/(1+\gamma)$ at the
present epoch. There is no much fine tuning in the numerical value
of $\gamma$ and dark energy density in this case gets naturally
connected to massive neutrino matter density at the present epoch. Note that we had 
introduced $\gamma$ for simplifying the analytical estimates
though we do not need to use it any further.
In our numerical elaboration we only consider $\alpha$ to be
the free parameter, since it controls $\rho_{\phi}/\rho_c$ in
matter and radiation dominated eras.

\begin{figure}
\centering
\includegraphics[width=0.45 \linewidth]{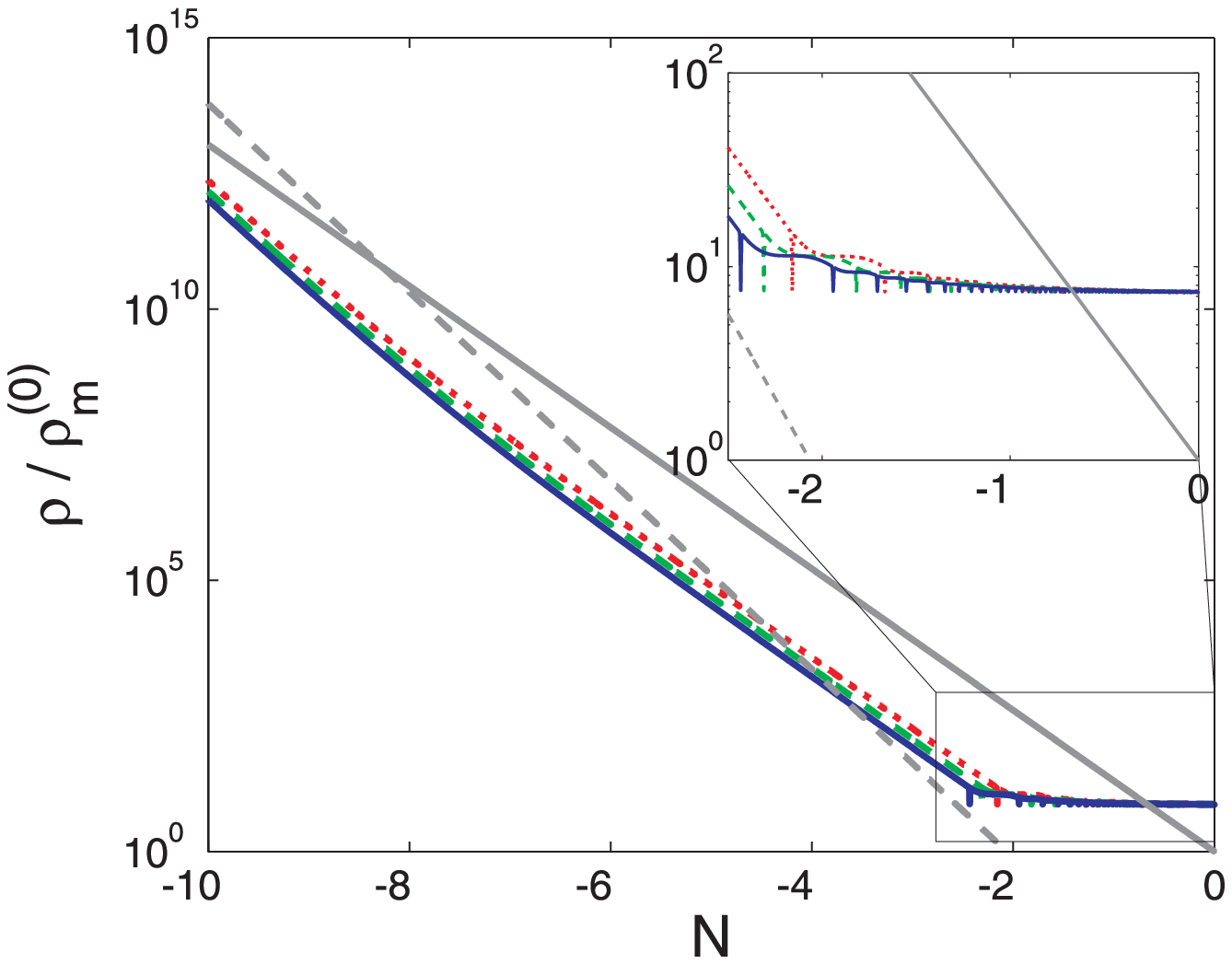}
\includegraphics[width=0.45 \linewidth]{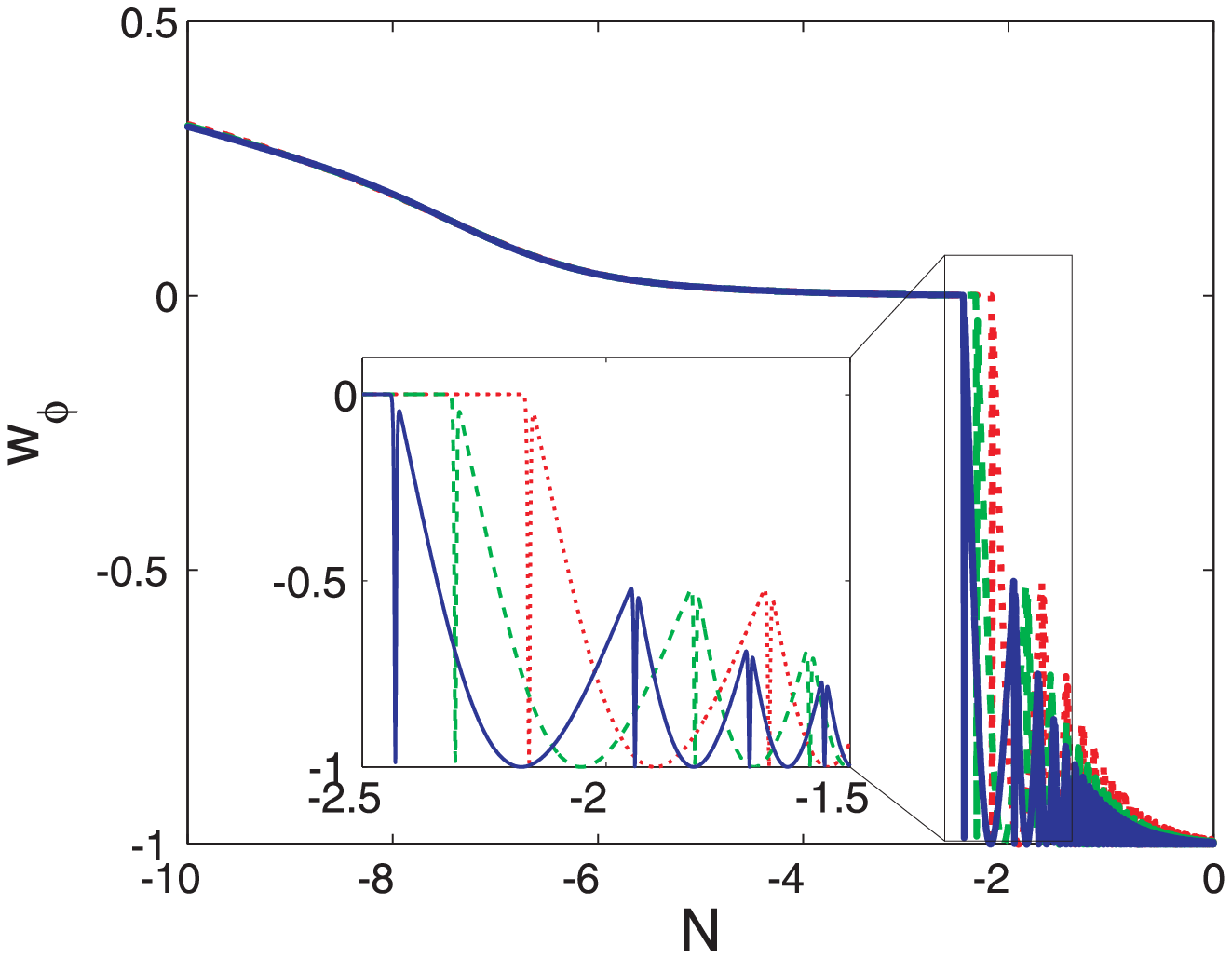}
\includegraphics[width=0.45 \linewidth]{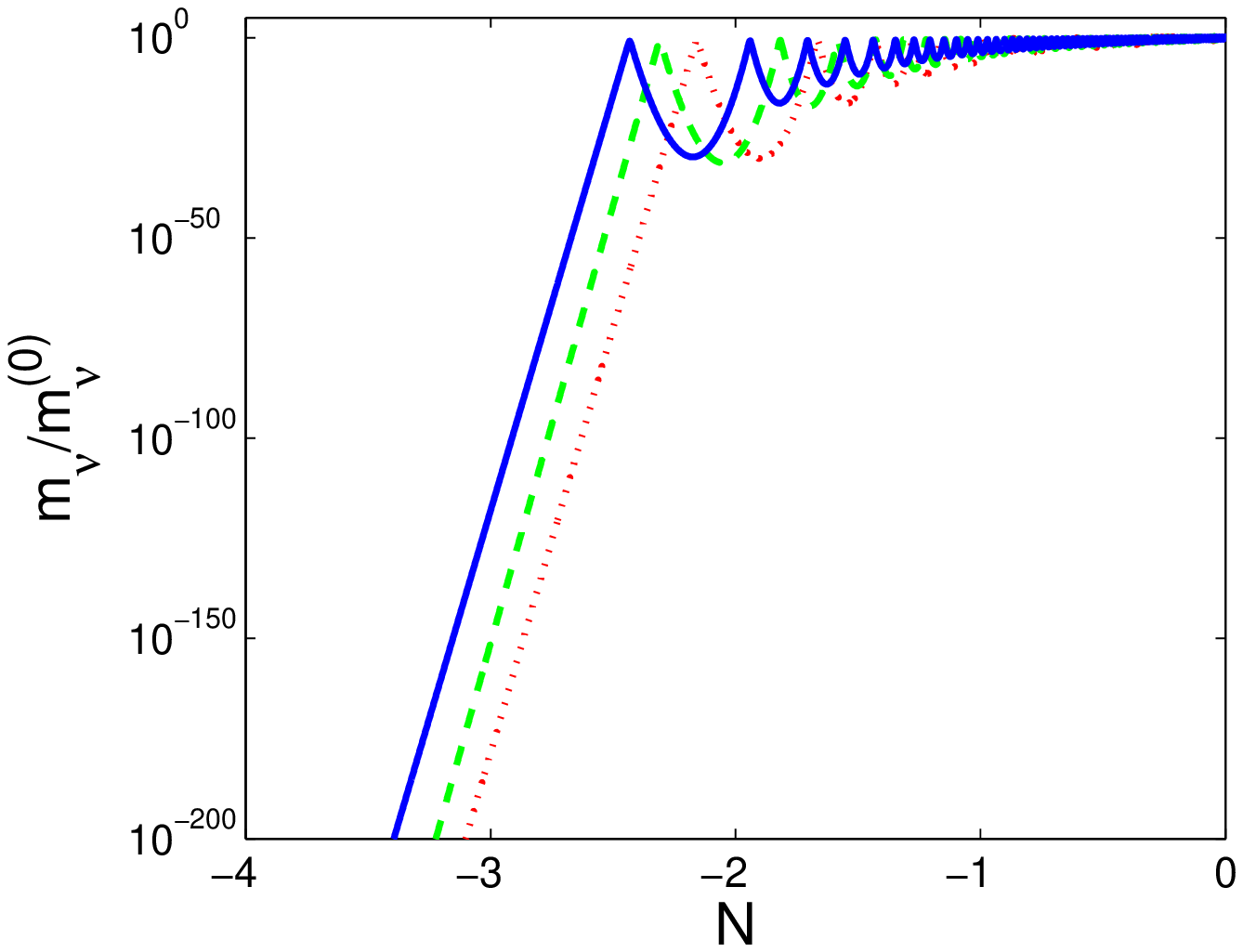}
\caption{{\it{Upper left: Evolutions of the various energy densities, normalized with the
matter energy density $\rho_{m}^{(0)}$ at present ($a_0=1$), as functions of $N \equiv
\ln a$, for the non-minimally coupled scenario, with the effective potential $V_{eff, 
\phi}
=-\alpha V_0 e^{-\alpha \phi/\Mpl}/\Mpl + \beta(\rho_{\nu} - 3 p_{\nu})/\Mpl$, where
$\rho_r$ is the gray-dashed curve, $\rho_m$ is the gray-solid curve, and
$\rho_{\phi}$ is plotted for three parameter choices, namely
$\alpha$= $30$ (blue-solid), $25$ (green-dashed) and $20$ (red-dotted),
with $V_0/\rho_{m}^{(0)}=7/3$.
Upper right: The corresponding evolution of the
quintessence equation-of-state parameter $w_\phi$. 
Lower: Evolutions of the neutrino masses relative to the current masses.
We have used $\Sigma m_{\nu} = 0.2$~eV,
$\Omega_m h^2 =0.118$ and $\rho_r^{(0)}/ \rho_m^{(0)}=2.58\times 10^{-4}$ as the boundary 
conditions.
}}}
\label{fig1}
\end{figure}

In order to examine the above behavior in more detail, we perform a numerical elaboration
and we depict the corresponding evolution of the
various energy densities,
the quintessence
equation-of-state parameter and the neutrino masses in Fig.~\ref{fig1}.
We should emphasize that thanks to the tracker behavior in the high
redshift regime and the emergence of the dark-energy dominated era at late
times, the scenario can successfully be embedded in the framework of
quintessential inflation (see Ref.~\citep{Hossain:2014xha,Hossain:2014zma} for details).
Additionally, since the early-time and late-time evolutions are insensitive to the 
initial 
conditions and
$V_0$, respectively, we choose $(\phi/\Mpl,~\dot{\phi}/\Mpl)=(-60/\alpha,0)$ at $N=-20$ 
and fix $V_0$ for convenience as $V_0 = 7\rho_m^{(0)} /3$, which corresponds to the dark 
energy scale. For 
this $V_0$ choice, the final scalar approaches to the origin of coordinate, i.e., $\phi 
\sim 0$ at 
$z=0$. Furthermore, we
should also mention that in this work we focus on the simplest
varying neutrino-mass model, with only one neutrino flavor having
the field varying mass, while the other two are treated as massless.

\begin{figure}
\centering
\includegraphics[width=0.45\linewidth]{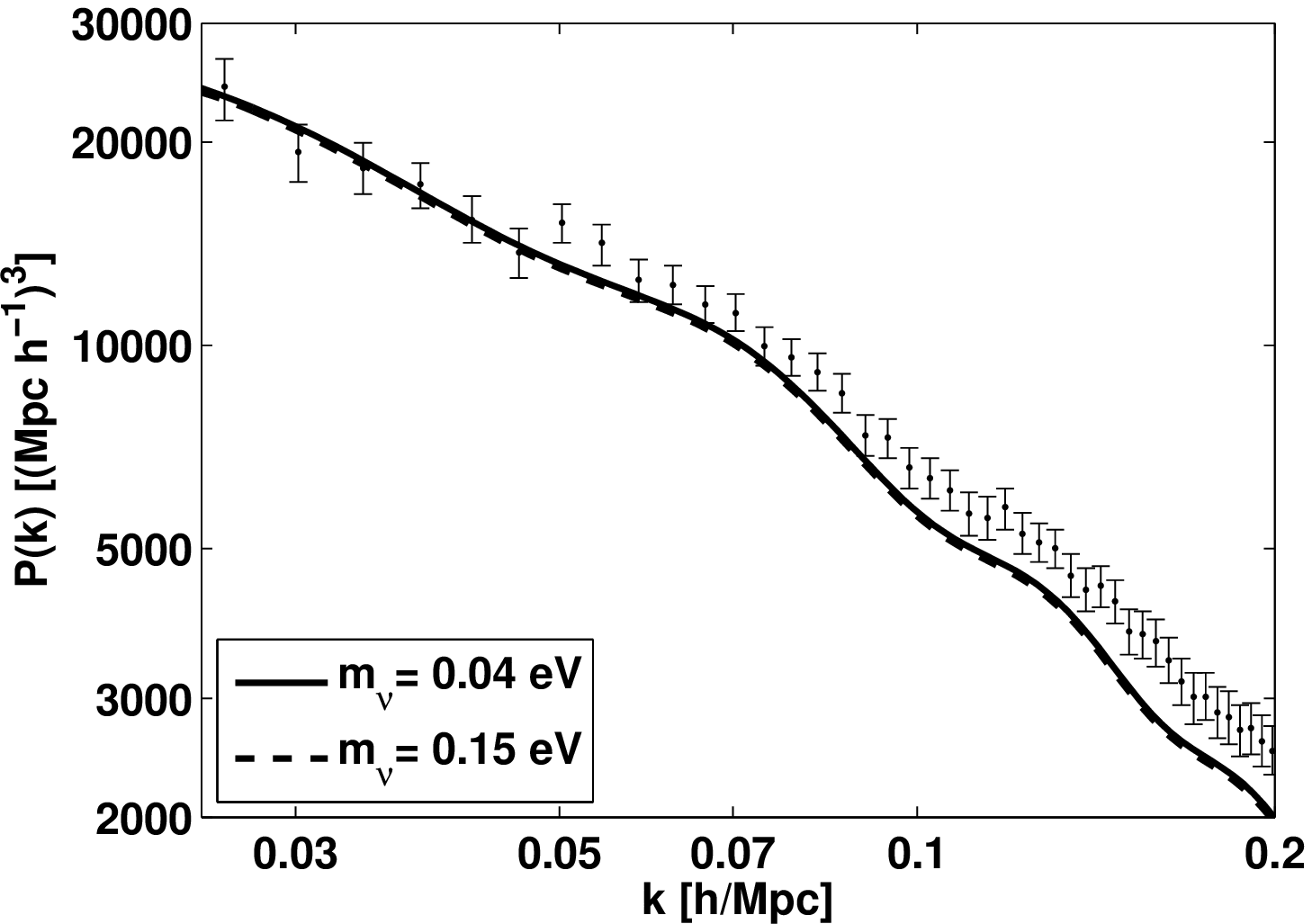}
\includegraphics[width=0.45 \linewidth]{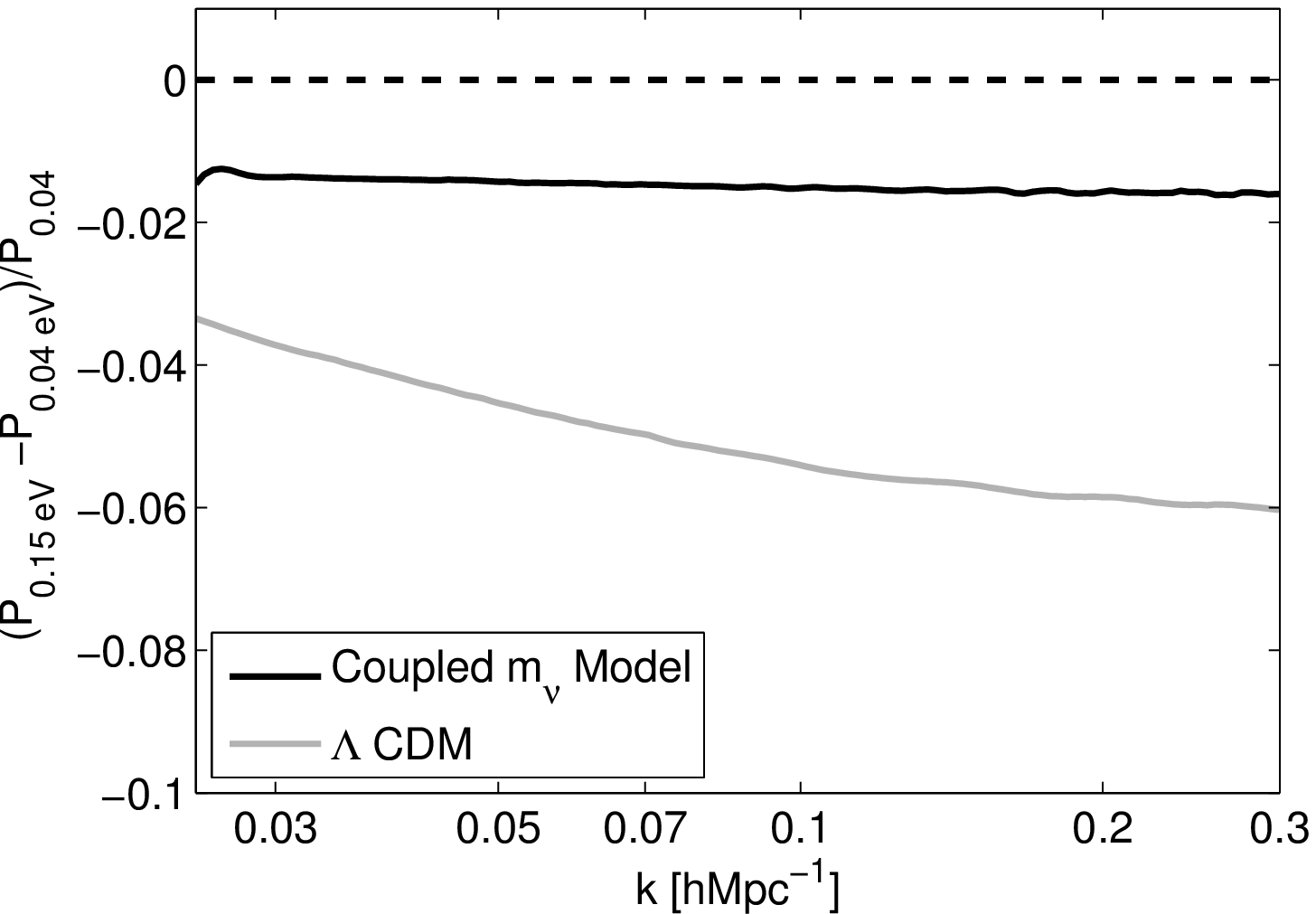}
\caption{ \it{Left: Matter power spectrum of the non-minimally coupled neutrino matter 
model with 
$V_{eff, \phi}
=-\alpha V_0 e^{-\alpha \phi/\Mpl}/\Mpl + \beta(\rho_{\nu} - 3 p_{\nu})/\Mpl$, where 
$\Sigma m_{\nu}
 = 0.04$~eV (solid) and $0.15$~eV (dashed),
$V_0/\rho_{m}^{(0)}=7/3$ as the boundary condition and ($\alpha, \Sigma 
m_{\nu}/\mathrm{eV}, \Omega_
ch^2, 100 \Omega_bh^2$)= $(25, 0.2, 0.118, 2.20)$.
Right: the deviation of the matter power spectrum $\Delta P^C \equiv \left( P^C_{0.15} - 
P^C_{0.04} 
\right) / P^C_{0.04} $ in coupled neutrino matter model (black solid line) and $\Delta 
P^{\Lambda 
CDM} \equiv \left( P^{\Lambda CDM}_{0.15} - P^{\Lambda CDM}_{0.04} \right) / P^{\Lambda 
CDM}_{0.04} 
$ in $\Lambda$CDM model (gray solid line), where $P_{0.15, 0.04} \equiv P(k)$ with the 
neutrino 
mass $\Sigma m_{\nu} = 0.15$ and $0.04$~eV.
}}
\label{figMPK}
\end{figure}

As the matter density perturbation is suppressed by the free-streaming massive neutrino,
the neutrino mass influences the matter power spectrum.
However, since small neutrino masses are preferred for  non-minimally coupled neutrino 
matter in the early time
(see also the lower panel in Fig.\ref{fig1}),
 the  suppression in the matter power spectrum due to massive neutrinos is minimized. In 
the 
upper 
panel of 
Fig.~\ref{figMPK},
 we depict the matter power spectrum with
$\Sigma m_{\nu} = 0.04$ and $0.15$~eV,
where the measured data points are from the Sloan Digital Sky
Survey (SDSS DR7).
In the lower panel of Fig. \ref{figMPK}, we plot the matter power spectrum deviation 
between $\Sigma m_{\nu}=0.15$~eV and $0.04$~eV in coupled neutrino matter (black solid 
line) and $\Lambda$CDM 
(gray solid line) models. Compared to $\Lambda$CDM, we can see that in the nonminimally 
coupled 
neutrino matter model the suppression of the matter power spectrum is minimized due to 
the free-streaming neutrino.

As shown in Refs.~\cite{Mota:2008nj, Ayaita:2014una}, the scalar $\phi$ within a large 
scale neutrino lump largely deviates from that at the background level $\bar{\phi}$, 
leading the neutrino mass to become negligible inside the neutrino lump, and thus
the perturbation to become non-linear.
In this work we desire to avoid such a non-linear region and we assume that the neutrino 
mass $m_{\nu}$ and the 
scalar $\phi$ are both homogeneous, and therefore the perturbation of neutrinos in the 
nonminimally-coupled 
scenario behaves as that in the $\Lambda$CDM paradigm:
\begin{eqnarray}
\dot{\delta}_{\nu} = 3 H \left( w_{\nu} - \frac{\delta p_{\nu}}{\delta \rho_{\nu}} \right) 
\delta_{\nu} - (1+w_{\nu})\left( \theta - \frac{\dot{h}}{2} \right) \,.
\end{eqnarray}
\begin{figure}
\centering
\includegraphics[width=.45 \linewidth]{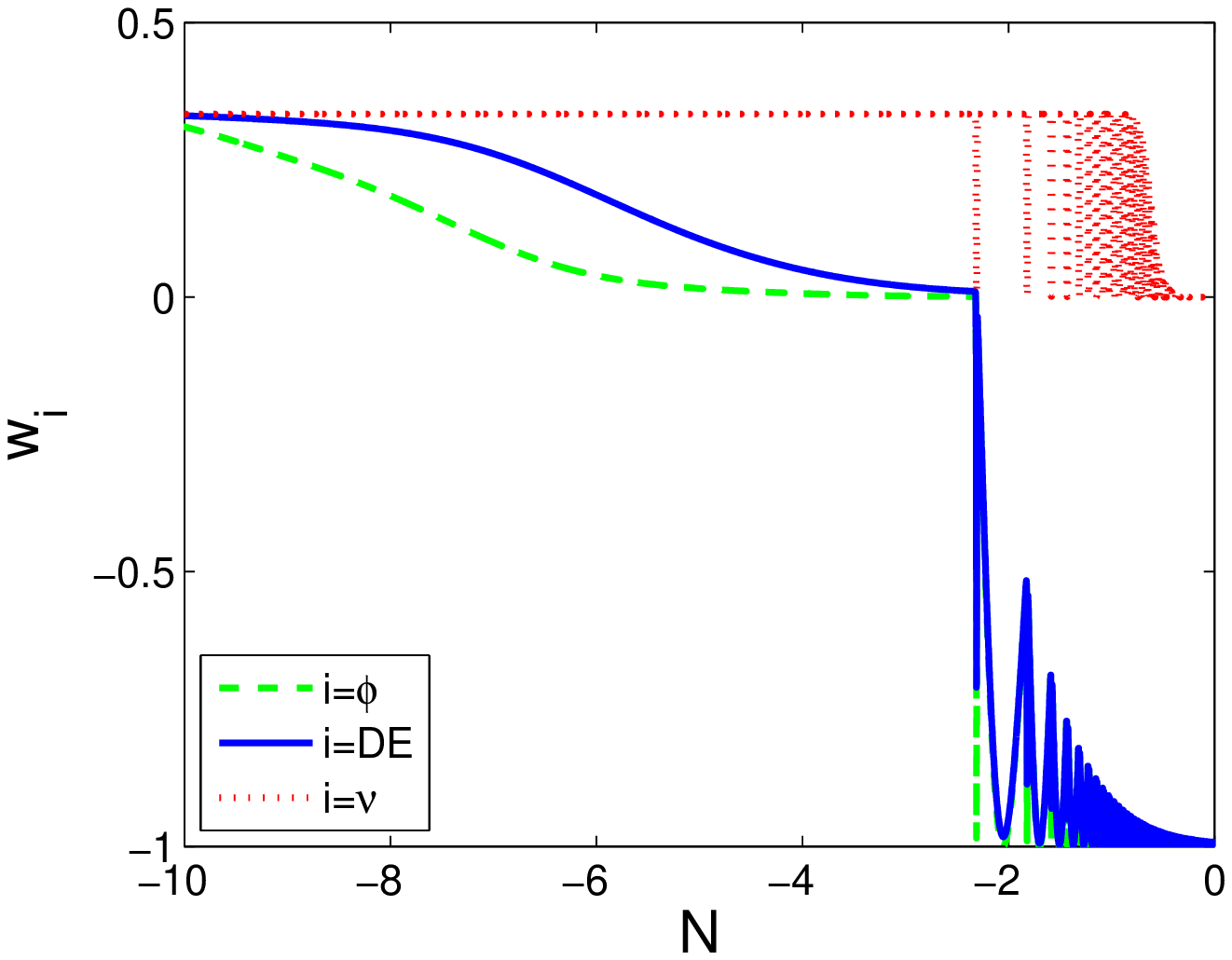}
\caption{{\it{Evolutions of the equation-of-state parameters of  the total dark-energy 
sector $w_{DE}$ (blue-solid), of the quintessence
field $w_\phi$ (green-dashed), and of the
neutrinos $w_\nu$ (red-dotted), as functions of $N \equiv
\ln a$, for the non-minimally coupled scenario, with the effective potential $V_{eff, 
\phi}
=-\alpha V_0 e^{-\alpha \phi/\Mpl}/\Mpl + \beta(\rho_{\nu} - 3 p_{\nu})/\Mpl$, where we 
have used
$\alpha$= $25$ and $V_0/\rho_{m}^{(0)}=7/3$ as well as $\Sigma m_{\nu} = 0.2$~eV,
$\Omega_m h^2 =0.118$ and $\rho_r^{(0)}/ \rho_m^{(0)}=2.58\times 10^{-4}$ as the boundary 
conditions
.}}}
\label{figwtot}
\end{figure}

Finally, since in the scenario at hand the dark energy sector is
attributed to the combination of the quintessence field and the
varying-mass neutrinos according to (\ref{rhoDE}) and (\ref{pDE}),
for completeness in Fig.~\ref{figwtot} we depict the evolution of
$w_\phi=p_\phi/\rho_\phi$ along with
$w_{DE}=(p_\phi+p_\nu)/(\rho_\phi+\rho_\nu)$ and
$w_\nu=p_\nu/\rho_\nu$. As expected according to our previous
discussion, at late times  $w_{\phi}$ determines $w_{DE}$ almost
completely. In addition, from (\ref{Veff}) we estimate
that
\begin{eqnarray}
\label{eq:est_beta}
\beta \simeq \frac{\alpha V(\phi_{min})}{\rho_{\nu}} \simeq \frac{32 \alpha}{\Sigma 
m_{\nu} 
/\mathrm{
eV}} \,,
\end{eqnarray}
where we have used $\rho_{\phi} \simeq V(\phi_{min})$, $\Omega_{\phi}h^2 = 0.34$ and 
$\Omega_{\nu} 
h^2 =
\Sigma m_{\nu}/ 94.1$~eV. Combined with (\ref{eq:mnu}), the small
fluctuation of $\phi$ results in a significant change in neutrino
masses. As shown in Fig.~\ref{figwtot},  neutrino masses along
with the equation-of-state parameters oscillate in $-2 \lesssim \ln
a \lesssim -0.5$, i.e. $0.7 \lesssim z \lesssim 6.4$, and become
purely massive after $z \lesssim 0.7$.

As the scenario at hand leads to a successful description of the
universe history at the background level, it is necessary to
confront it with observations-related matter perturbations, as the
varying-mass neutrino could in principle lead to strong constraints
in this case. In the following section we perform such an
observational analysis in detail.

\section{Observational constraints} \label{sec:4}

We use the program CosmoMC~\citep{Lewis:1999bs, Lewis:2002ah} to
extract the observational constraints on the scenario of the
previous section. An important point is that one should go beyond
the background evolution, since the massive neutrino effects will
also arise from the matter density perturbation. Therefore, in our
analysis we incorporate the matter power-spectrum data sets, which
include cosmic microwave background (CMB) from
Planck~\citep{Ade:2013zuv} and WMAP~\citep{Hinshaw:2012aka}, baryon
acoustic oscillation (BAO)  from Baryon Oscillation Spectroscopic
Survey (BOSS)~\citep{Anderson:2012sa, Anderson:2013zyy}, Type-Ia
supernova (SNIa) from Supernova Legacy Survey (SNLS)~\citep{Astier:2005qq}, and matter 
power 
spectrum from Sloan 
Digital
Sky Survey (SDSS DR4)~\citep{AdelmanMcCarthy:2005se} and WiggleZ Dark
Energy Survey~\citep{Blake:2011wn, Blake:2011en}. The details of the
fitting procedure can be found in Refs.~\citep{Lewis:1999bs, Lewis:2002ah}.
\begin{figure}
\centering
\includegraphics[width=0.45 \linewidth]{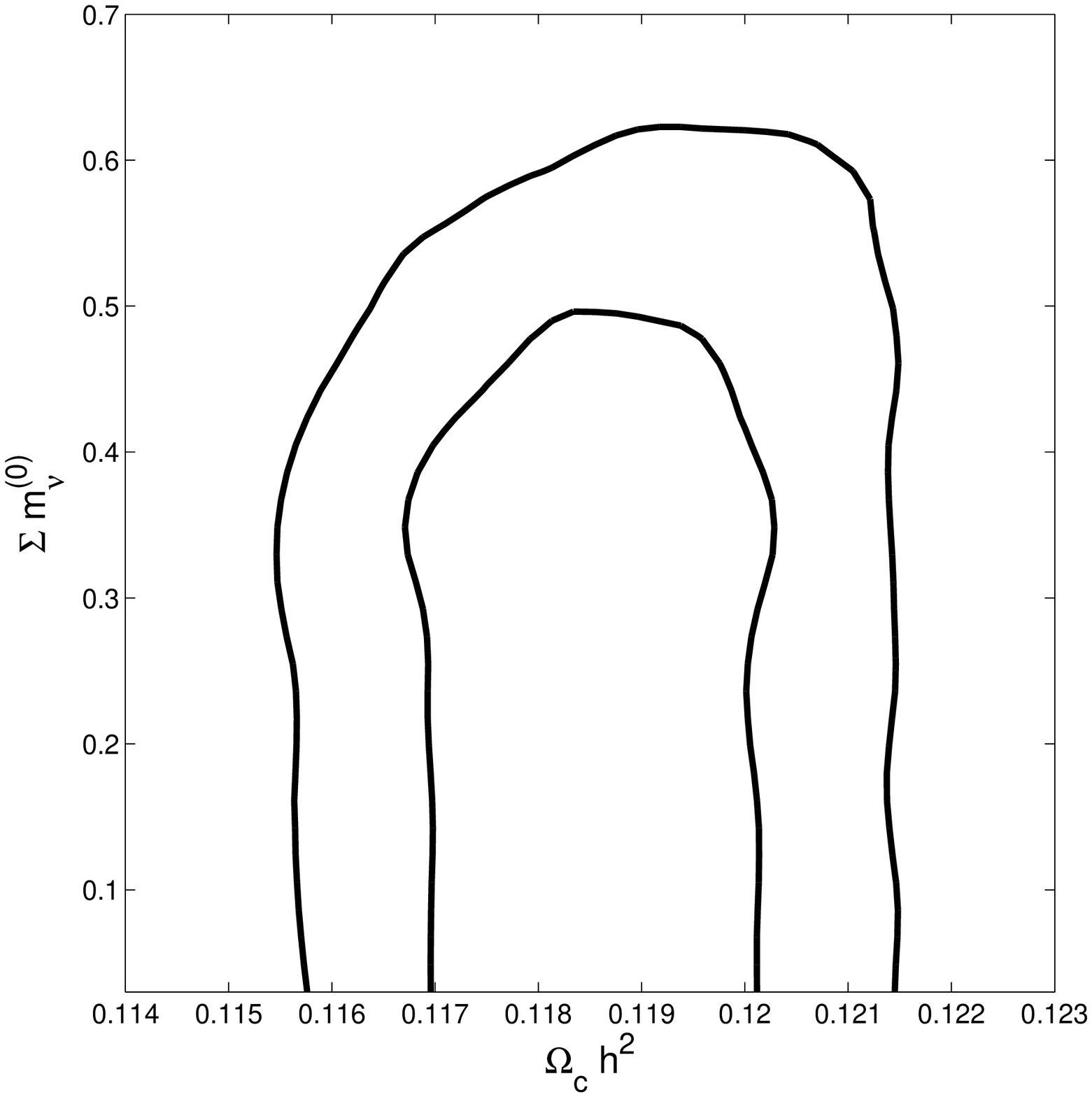}
\includegraphics[width=0.45 \linewidth]{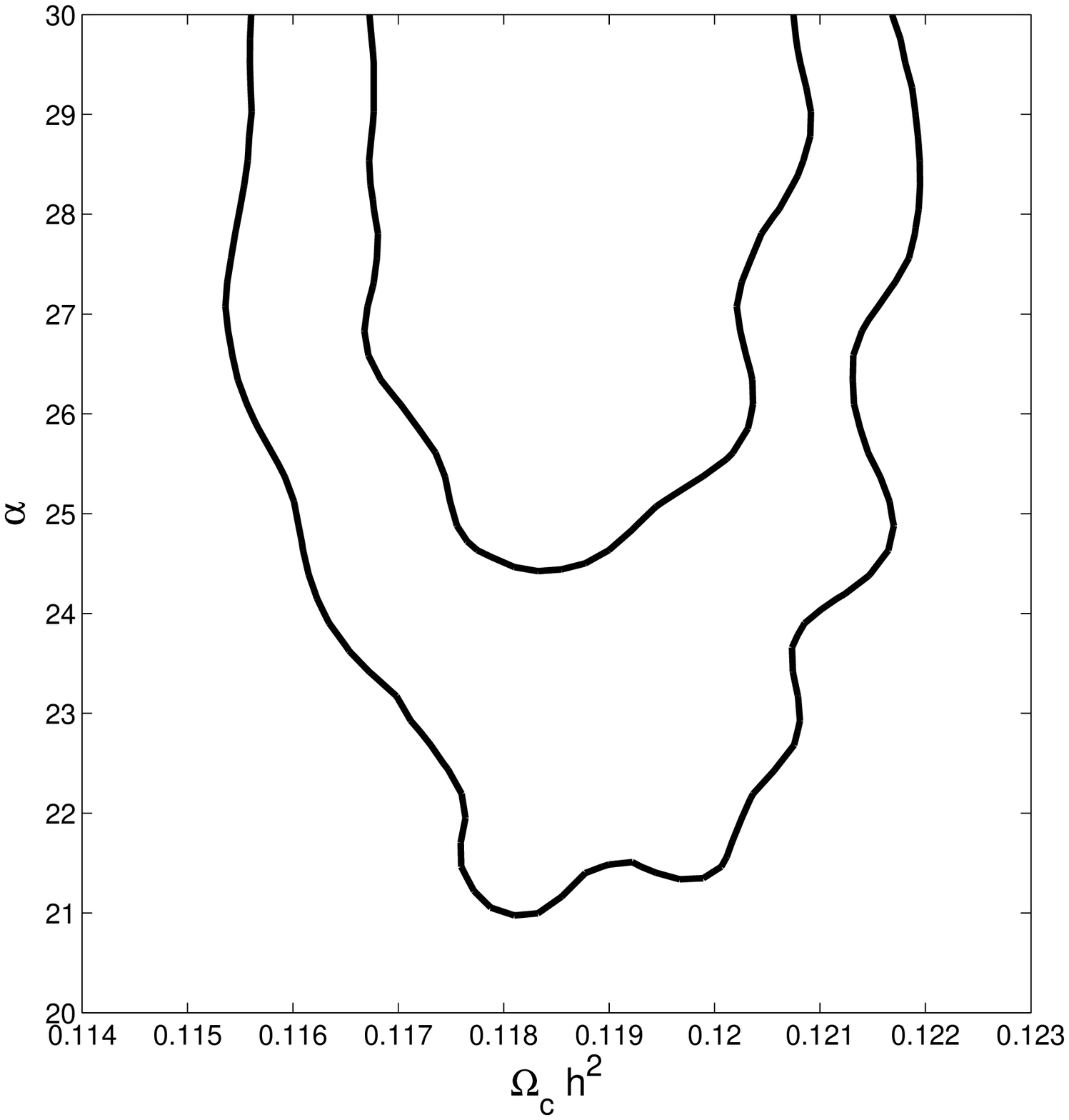}
\caption{{\it{Likelihood contours of the sum of the mass of the three neutrino species
$\Sigma m_{\nu}^{z=0}$ in eV (upper) and the potential parameter $\alpha$ (lower) versus 
the CDM 
physical density $\Omega_c h^2$, for the
non-minimally coupled scenario with the effective potential $V_{eff, \phi}
=-\alpha V_0 e^{-\alpha \phi/\Mpl}/\Mpl + \beta(\rho_{\nu} - 3 p_{\nu})/\Mpl$, where the 
inner and 
outer curves represent $1$ and $2\sigma$
confidence levels, respectively.
 }}}
\label{fig5a}
\end{figure}
In Fig.~\ref{fig5a}, we depict the $2D$ likelihood contours for the
mass sum of the three active neutrinos at present ($z=0$) and the model
parameter $\alpha$, versus the physical density of cold dark matter
(CDM) $\Omega_c h^2$. The scenario at hand is in agreement with
observations, and indeed quantities like the present dark-matter and
dark-energy density parameters have similar ranges as those in
$\Lambda$CDM cosmology~\citep{Ade:2013zuv}. However, the neutrino
mass sum in the model is enhanced to $0.28$~eV and the allowed window is
also significantly relaxed such that $\Sigma m_{\nu} < 0.52$~eV. The
potential parameter $\alpha$ controls the ratio $\rho_{\phi}/\rho_c$
in the high redshift regime, which at the $2-\sigma$ confidence level
is constrained to be $\alpha > 22.1$, which naturally yields the tracker behavior 
so that the coincident problem could be understood.
 Our results for the non-minimally coupled scenario are summarized in Table.~\ref{table1}.
Finally, it is worth to mention that the coupling $\beta$ is tuned to control 
$\Omega_{\phi}$ in 
our program, and its value is roughly inversely proportional to $\alpha$ and $\Sigma 
m_{\nu}$. In 
our numerical computation, we take $\beta \sim 800/\Sigma m_\nu$ with $\alpha=25$. This 
can be 
estimated by the relation of Eq.~(\ref{eq:est_beta}).
\begin{table*}
\begin{center}
\caption{List of priors for parameters and allowed region with $95\%$ C.L., and $V_{eff, 
\phi}
=-\alpha V_0 e^{-\alpha \phi/\Mpl}/\Mpl + \beta(\rho_{\nu} - 3 p_{\nu})/\Mpl$.  }
\begin{tabular}{|l||c|c|c|} \hline\hline
Parameter & Prior & Our result ($95\%$ C.L.) & $\Lambda$CDM ($95\%$ C.L.)
\\ \hline
Baryon density & $0.5 <100\Omega_bh^2<10$ & $ 100 \Omega_bh^2 =   2.19^{+0.04}_{-0.05}$ &
$ 100 \Omega_bh^2 =  2.22^{+0.04}_{-0.06}$
\\ \hline
CDM density & $10^{-3}<\Omega_ch^2<0.99$ & $ \Omega_ch^2 =  0.119 \pm 0.003$ & $
\Omega_ch^2
=  0.117^{+0.004}_{-0.002}$
\\ \hline
Neutrino mass & $0.03<\Sigma m_{\nu} < 1 $ eV & $\Sigma m_{\nu} = 
0.278^{+0.245}_{-0.248}$ 
eV
& $\Sigma m_{\nu} < 0.198$ eV
\\ \hline
Spectral index & $ 0.9 < n_s < 1.1$ & $ n_s = 0.963 ^{+0.010}_{-0.011}$ & $ n_s =
0.963^{+0.010}_{-
0.011} $
\\ \hline
Potential  & $ 20 < \alpha < 30$ & $\alpha > 22.1$ & $-$
\\ \hline
\end{tabular}
\label{table1}
\end{center}
\end{table*}

\section{Conclusions} \label{sec:conclusion}

We have examined the tracker quintessence models with inverse power-law and  
double-exponential potentials. In the first case, it is
difficult to obtain a viable scenario for generic values of model
parameters, whereas in the second case we are forced to make a
natural choice for the scale of the potential which brings back the
problem that cosmological constant is plagued with.
 With a hope to alleviate the problem, we have
considered massive neutrino matter non-minimally coupled to gravity.
During radiation and early matter era, massive neutrinos exhibit
relativistic behavior which implies vanishing coupling to scalar
field. Only at late times, when massive neutrinos turn
non-relativistic, their direct coupling to field builds up, leading
to appearance of a minimum in the effective potential of the field.
 In this picture, the
minimum of the effective potential is insensitive to the scale of
the potential $V_0$ and is rather given by the numerical value of $
\rho_\nu$ around the present epoch. At late times, when the scalar
field rolls around the minimum of the effective potential, the
dynamical system enters into scaling regime, which is an attractor.
Thanks to non-minimal coupling, the scaling solution is accelerating.
Clearly, the model under consideration has an edge over standard
quintessence with flat potential.

We have found that the neutrino mass grows in time and its mass in
the past is smaller than the current stage. In the early universe,
the free-streaming length of the neutrino around the horizon scale
is the same as that of the massless one. As shown in
Fig.~\ref{figwtot}, the neutrinos become massive only at the very
recent epoch. As a result, our scenario can relax the suppression of
the matter power spectrum from the free-streaming massive neutrino
in the early universe. The constraint on the neutrino masses is
reduced in comparison to  $\Lambda$CDM. In particular, in the scenario under
consideration, we found that the best-fit value of the neutrino
mass sum are around $0.28$~eV, along with the allowed window extended
to $0.52$~eV,
 which is consistent with particle physics experiments.

\section*{Acknowledgments}
 MS and ENS wish to thank National Center for Theoretical Sciences,
Hsinchu, Taiwan for the hospitality during the initial stages of
this work. This work was partially supported by National Center for
Theoretical Sciences,  National Science Council
(NSC-101-2112-M-007-006-MY3),
MoST (MoST-104-2112-M-007-003-MY3)  and National Tsing Hua
University~(104N2724E1).

\appendix

\section*{APPENDIX: THE INVERSE POWER-LAW POTENTIAL}

\begin{figure}
\centering
\includegraphics[width=0.45 \linewidth]{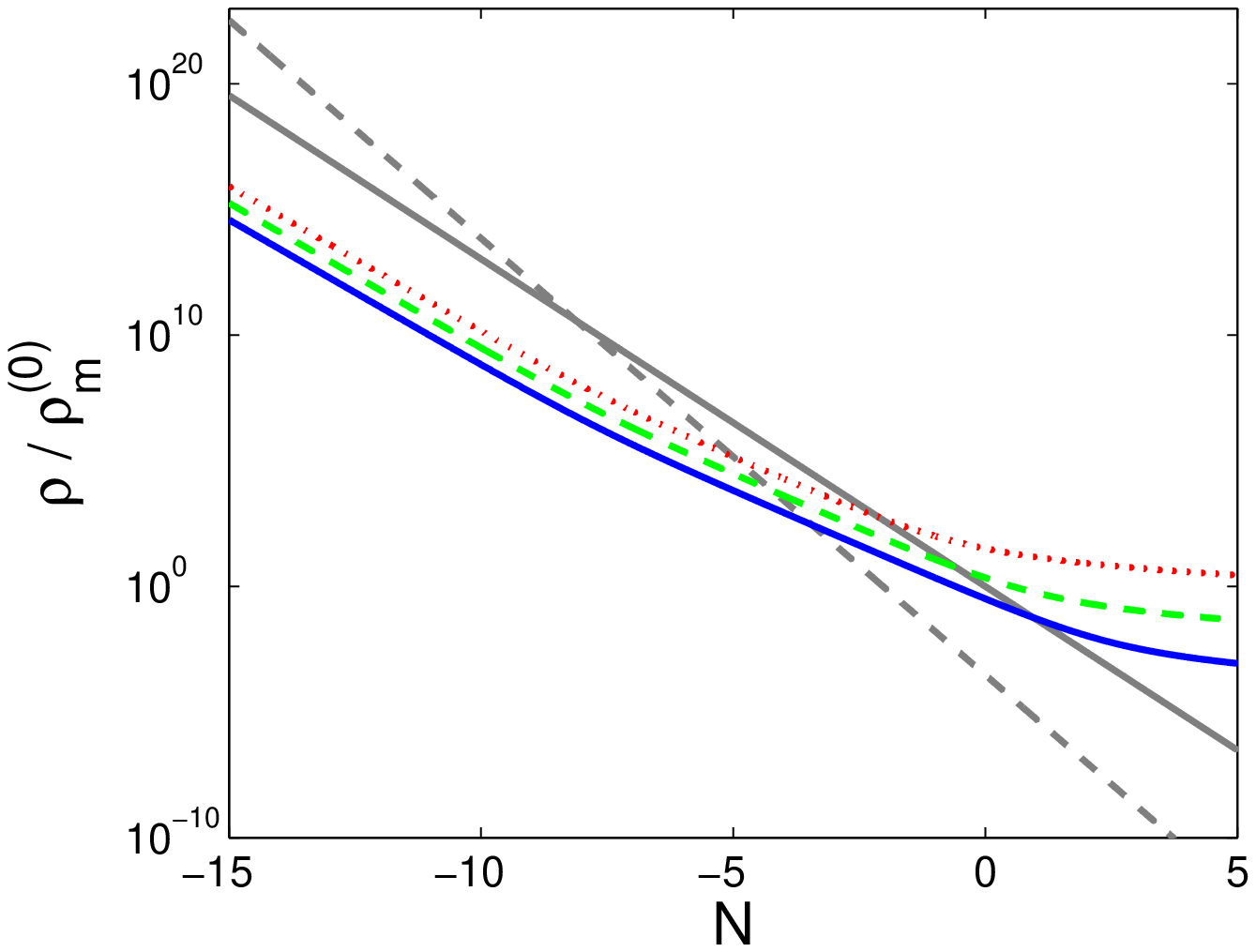}
\includegraphics[width=0.45 \linewidth]{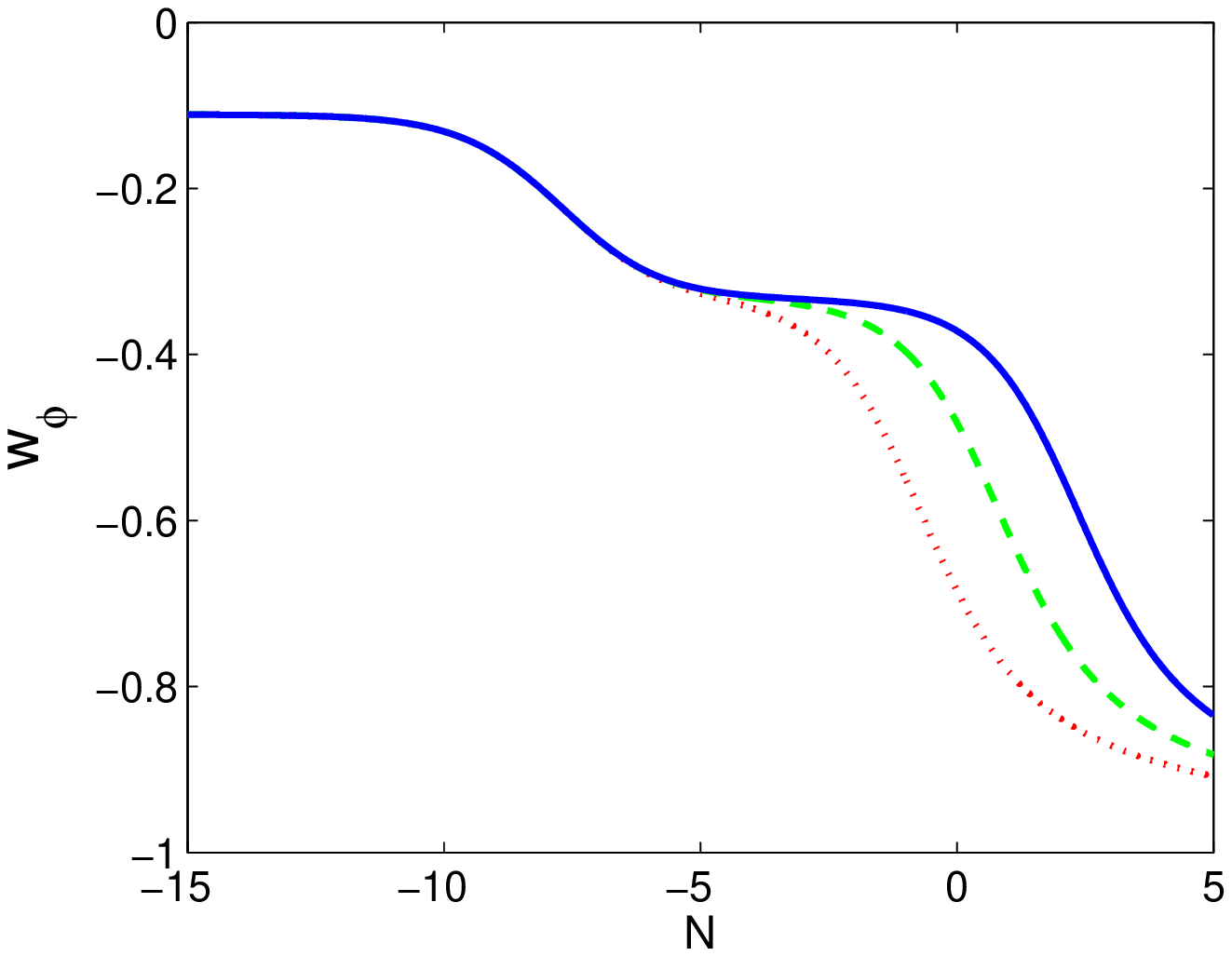}
\caption{{\it{Left: Evolutions of the various energy densities,
normalized with the matter energy density $\rho_{m}^{(0)}$ at
present ($a_0=1$), as a function of $N \equiv \ln a$, for the
minimally coupled scenario with the inverse power-law potential $V_0
\left( \frac{\phi}{\Mpl} \right)^{-n}$:
$\rho_r$ is the gray-dashed curve, $\rho_m$ is the gray-solid curve,
and $\rho_{\phi}$ is plotted for three parameter choices, namely
($n, V_0/\rho_{m}^{(0)}$)= $(4, 1)$ (blue-solid),  ($n,
V_0/\rho_{m}^{(0)}$)= $(4, 10^2)$ (green-dashed) and  ($n,
V_0/\rho_{m}^{(0)}$)= $(4, 10^4)$ (red-dotted). 
Right: The
corresponding evolution of the quintessence equation-of-state
parameter $w_\phi$. We have used $\rho_r^{(0)}/ \rho_m^{(0)} = 3
\times 10^{-4}$ as the boundary condition.}} } \label{fig2}
\end{figure}

For completeness, let us investigate the case of the power-law potential 
(\ref{eq:potphi4}). In this case, the evolution equations can be
explicitly integrated in the background (radiation/matter) dominated
regime. The  asymptotic behavior of the solution in the radiation
and matter dominated eras is given by a power-law behavior with
\begin{eqnarray}
\label{eq:eq:w_phi2} w_{\phi} = \frac{-2+n w_b}{n+2} \,.
\end{eqnarray}
 Eq.~(\ref{eq:eq:w_phi2}) implies that $w_{\phi} < w_b$ for $n>0$ in the radiation and 
matter 
dominated
epochs, i.e. $\rho_{\phi}/\rho_m^{(0)}$ increases, allowing finally
to catch up with dark-energy domination. For moderate values of
$n\sim 1$, the field might take over the background and as a result
$\rho_\phi$ might obtain the observed value of dark energy at the
present epoch, whereas $\omega_\phi$ has not yet reached the desired
value. In the upper graph of Fig.~\ref{fig2}, we depict the
corresponding evolutions for the energy densities of radiation
($\rho_r$), matter ($\rho_m$) and quintessence field
($\rho_{\phi}$), normalized with the matter energy density
$\rho_{m}^{(0)}$ at present ($a_0=1$), as functions of $N \equiv \ln
a$, for the case $n=4$ with three choices of $V_0$. In the lower
graph of Fig.~\ref{fig2}, we show the corresponding evolution for
the quintessence equation-of-state parameter $w_\phi$. From this
plot we can see that there exist two regimes with constant values of
$w_{\phi} \approx-1/9$ and $w_{\phi}\approx -1/3$, in the radiation
and matter dominated era, where the field tracks the background. Clearly, the 
equation-of-state parameter $w_\phi$ does not reach the required
value at the present epoch.

The above situation can be remedied by taking large values of $n$, for which
$w_{\phi} \to w_b$. In this case, the inverse power-law potentials
reduce to the exponential form, leading to scaling solution. The late-time exit from 
the scaling regime is guaranteed by the shallow nature of
inverse power-law functions, though we need to tune the model
parameters appropriately. Firstly, irrespectively of the quintessence
model, using the slow-roll parameter $\eta$ we find that
$m_\phi\sim H_0\sim 10^{-33}$ eV. The slow-roll parameter $\epsilon$
tells us that this would happen at the present epoch, provided that
$\phi_0/\Mpl>>\sqrt{2}n$. Hence, this implies that $V_0\sim n^n
\rho_{cr}$. In the case where $V_0\sim M_p$, one requires a large value of $n$. On the 
other hand, for moderate values of $n$ we need to choose $V_0\sim
\rho_{cr}$ (Sahni, Sami  \&   Souradeep  \citep{Sahni:2001qp}).

\end{document}